\documentstyle [12pt]{article}
\advance\voffset by -1 in
\advance\hoffset by -.5 in
\begin{document}
\def\l{\lambda}
\def\p{\partial}
\def\pr{\prime}
\def\ti{\tilde}
\def\t{\tau}
\def\o{\omega}
\def\O{\Omega}
\def\d{\delta}
\def\hw{\hat{w}}
\def\w{\wedge}
\def\({\left(}
\def\){\right)}
\def\cA{{\cal A}}
\def\cB{{\cal B}}
\def\tr{{\rm tr}\:}
\centerline{{\large Poisson brackets with divergence terms in field theories:}}
\centerline{{\large two examples}}

\vspace{.3in}
\centerline{{\bf L. A. Dickey}}

\vspace{.3in}
\centerline{Department of Mathematics, University of Oklahoma\footnote
{e-mail: ldickey@ou.edu}}

\vspace{.5in}
\begin{abstract}

In field theories one often works with the functionals which are
integrals of some densities. These densities are defined up to
divergence terms (boundary terms). A Poisson bracket of two functionals is 
also a functional, i.e., an integral of a density. Suppose the divergence 
term in the density of the Poisson bracket be fixed so that it becomes a
bilinear form of densities of two functionals.
Then the left-hand side of the Jacobi identity
written in terms of densities is not necessarily zero but a divergence of a
trilinear form. The question is: what can be said about this trilinear
form, what kind of a higher Jacobi identity (involving four fields) it
enjoys? Two examples whose origin is the theory of integrable systems
are given.
\end{abstract}

In field theories one often works with the functionals which are
integrals of some densities. These densities are defined up to
divergence terms (boundary terms). A Poisson bracket of two functionals is also a
functional, i.e., an integral of a density. Suppose the divergence term 
in the density of the Poisson bracket be fixed so that it becomes a
bilinear form of densities of two functionals.
Then the left-hand side of the Jacobi identity
written in terms of densities is not necessarily zero but a divergence of a
trilinear form. The question is: what can be said about this trilinear
form, what kind of a higher Jacobi identity (involving four fields) it
enjoys?
Our examples relate to the simplest 1-dimensional case. Their origin is
in the theory of integrable systems.

My attention to this topic was called by J. Stasheff, to whom I am very
thankful. In a series of articles by V. O. Soloviev [1] a close problem was
posed: is it possible, using the freedom of choice of the divergence term in 
a local Poisson bracket, to make the Jacobi identity exact?\\

{\bf 1. Scalar example.}

We have the following structures.

1) Differential algebra $\cA$ consisting of differential polynomials of $u$ 
with the derivation $\p=d/dx$.

2) A space $\cB=\cA/\p\cA$. 

3) Derivations $\p_a=\sum_{i=0}^\infty a^{(i)}\p/\p u^{(i)}$ where $a\in\cA$. 
They commute with $\p$ and, therefore, can be transferred to $\cB$.
They form a Lie algebra with respect of the commutator $[\p_a,\p_b]=\p_a\p_b-
\p_b\p_a=\p_{\p_ab-\p_ba}$ called the Lie algebra of vector fields.

4) Vector fields $\p_{X'},$ where $X\in\cA$, form a subalgebra,
$$[\p_{X'},\p_{Y'}]=\p_{(\p_{X'}Y-\p_{Y'}X)'}.\eqno{(1.1)}$$

5) We define a skew symmetric bilinear form on the above subalgebra:
$$\o(\p_{X'},\p_{Y'})=(X'Y-XY')/2.\eqno{(1.2)}$$

{\bf Lemma 1.1.} {\sl The form $\o$ is closed with respect to the
derivation $\d$, acting as} $$(\d\o)(\p_{X'},\p_{Y'},\p_{Z'})=
\p_{X'}\o(\p_{Y'},\p_{Z'})-\o([\p_{X'},\p_{Y'}],\p_{Z'})+({\rm
cyclic}).$$ 

{\em Proof}. $$\p_{X'}2\o(\p_{Y'},\p_{Z'})-2\o([\p_{X'},\p_{Y'}],\p_{Z'})+({\rm
cyclic})$$ $$=\p_{X'}(Y'Z-YZ')-(\p_{X'}Y'-\p_{Y'}X')Z+(\p_{X'}Y-
\p_{Y'}X)Z'+({\rm cyclic})$$ $$=Y'\p_{X'}Z-Y\p_{X'}Z'+Z\p_{Y'}X'-
Z'\p_{Y'}X+({\rm cyclic})=0.~\Box$$Notice that this is an exact
equality, i.e., in $\cA$.

6) Now, we define a {\em quasi}-Poisson bracket in $\cA$: $$\{f,g\}=
\o(\p_{X'},\p_{Y'})=(X'Y-XY')/2,~{\rm where}~X=\d f/\d u,~Y=\d g/\d u.$$
The word {\em quasi} refers to the fact that the Jacobi identity 
holds, as we will see, only up to an exact derivative, i.e., in $\cB$.
This bracket is also well defined in $\cB$ since a variational
derivative of an exact derivative is always zero. Thus, considered in
$\cB$, this bracket is a genuine Poisson bracket\footnote{The Leibniz property
of a Poisson bracket is not required and even does not make any sense since there
is no multiplication in $\cB$.}.

Introduce a notation $\xi_f=\p_{X'}$ where $X=\d f/\d u$. The Poisson
bracket is a pull-back of the ``symplectic" form $\o$ under the mapping
$f\mapsto\xi_f$.

7) Integration by parts yields a formula
$$ \p_{X'}g=X'Y+\p\:\O_g(\p_{X'})\eqno{(1.3)}$$ where $Y=\d g/\d u$ and $\O_g$ is 
a 1-form linearly depending on $g$ (the Poincar\'e invariant for the Lagrangian
$g$; in terms of canonical variables ${\bf p}d{\bf q}$); $X$ is not
necessarily a variational derivative. If it is, then we write $\p_{X'}=\xi_f$. 

{\bf Lemma 1.2.}  $$\xi_fg-
\xi_gf=2\{f,g\}+\p\:\Phi(f,g)~{\rm where}~\Phi(f,g)=\O_g(\xi_f)-
\O_f(\xi_g).$$This is obvious. 

$\Phi$ is a skew symmetric bilinear form
in $\cA$ which cannot be transferred to $\cB$ since $\Phi(f,g')\neq 0$.

{\bf Lemma 1.3.}
$$\xi_fg=\Phi(f,g').$$ 

{\em Proof.} Take a derivative of Eq.(1.3): $\xi_fg'=\p(X'Y+\p\:\O_g(\xi_f
))$. On the other hand, this is $\p\:\O_{g'}(\xi_f)$ (from the same equation
taking into account that the variational derivative of $g'$ vanishes). Thus,
$\O_{g'}(\xi_f)=X'Y+\p\:\O_{g}(\xi_f)=\xi_fg$. Further, $\xi_{g'}=0$, hence
$\Phi(f,g')=\O_{g'}(\xi_f)=\xi_fg$.

{\bf Proposition 1.1.} $$\xi_{\{f,g\}}=[\xi_f,\xi_g].$$

{\em Proof.} For any vector field $\xi=\p_{Z'}$ we have $$0=(\d\o)(\xi_f,\xi_g,
\xi)=\xi_f\o(\xi_g,\xi)-\xi_g\o(\xi_f,\xi)+\xi\o(\xi_f,\xi_g)$$
$$-\o([\xi_f,\xi_g],\xi)+\o([\xi_f,\xi],\xi_g)-\o([\xi_g,\xi],\xi_f).$$
Using (1.3), we have $$\o(\xi_f,\xi)=(X'Z-XZ')/2=-XZ'+(XZ)'/2=-\xi
f+\p(\O_f(\xi)+(XZ)/2),\eqno{(1.4)}$$therefore, $$0=-\xi_f\xi g+\p(\xi_f\O_g(\xi)
+\xi_f(YZ)/2)+\xi_g\xi f-\p\:(\xi_g\O_f(\xi)+\xi_g(XZ)/2)$$ $$+\xi\{f,g\}-
\o([\xi_f,\xi_g],\xi)+[\xi_f,\xi]g-
\p\:(\O_g([\xi_f,\xi])+Y(\p_{X'}Z-\p_{Z'}X)/2)$$ $$-
[\xi_g,\xi]f+\p\:(\O_f([\xi_g,\xi])+X(\p_{Y'}Z-\p_{Z'}Y)/2)$$ $$=-
\xi(\xi_fg-\xi_gf)+\xi\{f,g\}-\o([\xi_f,\xi_g],\xi)
+\p\:(\xi_f\O_g(\xi)-\xi_g\O_f(\xi)-\O_g([\xi_f,\xi])+\O_f([\xi_g,\xi]))$$ 
$$+\p\:(\p_{X'}(YZ)-\p_{Y'}(XZ)-Y\p_{X'}Z+Y\p_{Z'}X+X\p_{Y'}Z-X\p_{Z'}Y)/2.$$
Lemma 1.2 implies: $$-\xi(\xi_fg-\xi_gf)+\xi\{f,g\}=-\xi\{f,g\}+\p\:\xi(-
\O_g(\xi_f)+\O_f(\xi_g)).$$
Replacing $f$ by $\{f,g\}$ in (1.4), we have
$$-\xi\{f,g\}=\o(\xi_{\{f,g\}},\xi)-\p\:(\O_{\{f,g\}}(\xi)+TZ/2)\eqno{(1.5)}$$ where
$\xi_{\{f,g\}}=\p_{T'}$.

First of all, we notice now that $\o(\xi_{\{f,g\}},\xi)-\o([\xi_f,\xi_g],\xi)$
is an exact derivative. The vector $\xi_{\{f,g\}}-[\xi_f,\xi_g]$ has a
form $\p_{U'}$. An expression of the type $UZ'-U'Z$ is an exact
derivative for an arbitrary $Z$ iff $U=$const and $\p_{U'}=0$. 
We arrive at the statement of the proposition. $\Box$

We actually did not use yet the exact form of terms with $\p$. They will
be important in what follows. Therefore, we must collect all remaining terms.
Notice that Proposition 1.1 implies that $T=\p_{X'}Y-\p_{Y'}X$. These
terms cancel with two other terms. The rest of terms are
$$\p\:(-\xi\O_g(\xi_f)+\xi\O_f(\xi_g)-\O_{\{f,g\}}(\xi)
+\xi_f\O_g(\xi)-\xi_g\O_f(\xi)-\O_g([\xi_f,\xi])+\O_f([\xi_g,\xi])$$
$$+(Y\p_{Z'}X-X\p_{Z'}Y)/2)=0.\eqno{(1.6)}$$ 

{\bf Proposition 1.2.} {\sl The Poisson bracket $\{f,g\}$ satisfies the Jacobi
identity up to an exact derivative term: $$\{\{f,g\},h\}+({\rm cyclic})=
\p\:\Psi(f,g,h)$$ where $\Psi=\d\Phi$ is a trilinear form,  $\Phi$ 
was defined earlier, and} $\d$: $$(\d\Phi)(f,g,h)=\xi_f
\Phi(g,h)-\Phi(\{f,g\},h)+({\rm cyclic}).$$ 

{\em Proof}. Let $X=\d f/\d u,~Y=\d g/\d u,~Z=\d h/\d u$. We have (using 
Proposition 1.1)
$$0=(d\o)(\xi_f,\xi_g,\xi_h)=\xi_f\{g,h\}-\o(\xi_{\{f,g\}},\xi_h)+({\rm
cyclic}).$$Transform this with the help of (1.5): $$0=-2\o(\xi_{\{g,h\}},\xi_f)
+\p\:(\O_{\{g,h\}}(\xi_f)+(\p_{Y'}Z-\p_{Z'}Y)X)+({\rm cyclic}).$$In the right-hand 
side one can perform any cyclic permutation of $X,Y$ and $Z$ in every term.
The equation can be rewritten as $$2\{\{f,g\},h\}+({\rm cyclic})=  
\p\:(\O_{\{f,g\}}(\xi_h)+(Y\p_{Z'}X-X\p_{Z'}Y)/2)+({\rm cyclic}).$$The last 
term can be eliminated with the help of (1.6):$$2\{\{f,g\},h\}+({\rm cyclic})=
\p\:(\O_{\{f,g\}}(\xi_h)+\xi_h\O_g(\xi_f)-\xi_h\O_f(\xi_g)$$ $$+\O_{\{f,g\}}
(\xi_h)-\xi_f\O_g(\xi_h)+\xi_g\O_f(\xi_h)+\O_g([\xi_f,\xi_h])-\O_f([\xi_g,
\xi_h])+({\rm cyclic}))$$ $$=2\p\:(\xi_h\O_g(\xi_f)-\xi_h\O_f(\xi_g)+\O_{\{f,g
\}}(\xi_h)-\O_h([\xi_f,\xi_g])+({\rm cyclic}))$$ $$=2\p\:(\xi_h\Phi(f,g)-
\Phi(\{f,g\},h)+({\rm cyclic}))=2\p\d\Phi(f,g,h).~\Box$$

{\bf Proposition 1.3.} {\sl The 3-form $\Psi$ satisfies the identity (a
``higher Jacobi'')
$$\sum_{i<j}(-
1)^{i+j}\Psi(\{f_i,f_j\},...,\hat{f}_i,...,\hat{f}_j,...)=0$$ for arbitrary 
4 functions $f_1,f_2,f_3$ and $f_4$.}

{\em Proof.} The derivation $\d$ is a differential (i.e., $\d^2=0$) only
in $\cB$, since in $\cA$ the Jacobi identity holds not precisely but up to a 
term. Therefore, the fact that $\Psi=\d\Phi$ does not imply $\d\Psi=0$. (The
proclaimed statement involves a part of the expression for $\d\Psi$).
Nevertheless, we can keep $\d^2$ under control.

It is not very difficult to compute that $$(\d^2\Phi)(f_1,f_2,f_3,f_4)=\sum_
{i<j<k}(-1)^l\Phi(\{\{f_i,f_j\},f_k\}+({\rm cyclic}),f_l),~~{\rm where}~l\neq 
i,j,k,$$ the only properties of the Poisson bracket used here are
bilinearity, skew symmetry and Proposition 1.1.
The first argument of $\Phi$ is not zero, it is
$\p\:\Psi(f_i,f_j,f_k)$. Thus, $$\d\Psi(f_1,f_2,f_3,f_4)=\sum_{i<j<k}(-1)^l
\Phi(\p\:\Psi(f_i,f_j,f_k),f_l)=\sum_{i<j<k}(-1)^{l-1}\Phi(f_l,\p\:\Psi(f_i,f_j,
f_k)).$$
Using lemma 1.3, one gets $$\d\Psi(f_1,f_2,f_3,f_4)=\sum_{i<j<k}(-1)^
{l-1}\xi_{f_l}\Psi(f_i,f_j,f_k)=\sum_l(-1)^{l-1}\xi_{f_l}\Psi(...,\hat{f_l},...).$$ 
On the other hand, by definition of the derivation $\d$, $$\d\Psi(f_1,f_2,f_3
,f_4)=\sum_l(-1)^{l-1}\xi_{f_l}\Psi(...,\hat{f_l},...)+
\sum_{i<j}(-1)^{i+j}\Psi(\{f_i,f_j\},...,\hat{f}_i,...,\hat{f}_j,...)$$
Equating two expressions, we obtain the required statement. $\Box$\\

{\bf 2. Matrix example.}

1) Elements $u_{jk}$ of an $n\times n$ matrix $U$ are taken as
generators of a differential algebra $\cA$.

2) Let $\cB=\cA/\p\cA$.

3) If $a$ is a matrix with entries belonging to $\cA$ then a derivation
in $\cA$ (a ``vector field'') can be defined: $$\p_a=\sum_{k,ij}a_{ij}^{(k)}{\p\over\p u_{ij}^{(k)}}
=\tr a^{(k)}{\p\over\p U^{(k)}},~~{\rm where}~\left({\p\over\p
U^{(k)}}\right)_{ij}={\p\over\p u_{ji}^{(k)}}.$$

4) To any matrix $X$ (hereafter we always assume that all scalars,
elements of matrix, etc always belong to $\cA$) another matrix is
assigned, $H(X)=X'+[U,X]$ and also a vector field $\p_{H(X)}$.

It is easy to check that $$ [\p_{H(X)},\p_{H(Y)}]=\p_{H([X,Y]+\p_{H(X)}Y-
\p_{H(Y)}X)}. \eqno{(2.1)}$$ This, in particular, means that the vector
fields of a special type $\p_{H(X)}$ make a subalgebra.

5) Define a skew symmetric bilinear form on the above subalgebra:
$$\o(\p_{H(X)},\p_{H(Y)})=\tr(H(X)Y-H(Y)X)/2.\eqno{(2.2)}$$

{\bf Lemma 2.1.} {\sl The relation $$\d\o(\p_{H(X)},\p_{H(Y)},\p_{H(Z)})
=0$$ holds, i.e., the form is closed exactly, in $\cA$.}

The lemma can be verified by a direct calculation.

6) The Poisson bracket is $$\{f,g\}=\o(\xi_f,\xi_g),~{\rm
where}~\xi_f=\p_{H(X)},~\xi_g=\p_{H(Y)},~{\rm and}~X=\d f/\d U,~Y=\d g/\d U
.$$ Here, $\d f/\d U$ is a matrix with the entries: $(\d f/\d U)_{ij}=
\d f/\d u_{ji}$.

7) The analogue to the formula (1.3) is now $$\p_{H(X)}g=\tr H(X)Y+\p\:\O_g
(\p_{H(X)}) \eqno{(2.3)}$$ where $\O_g$ is a 1-form, $X$ is an arbitrary
matrix, $Y=\d g/\d U$. An obvious corollary is:

{\bf Lemma 2.2.} {\sl The formula 
$$\xi_fg-\xi_gf=2\{f,g\}+\p\:\Phi(f,g)~{\rm where}~\Phi(f,g)=\O_g(\xi_f)-
\O_f(\xi_g)$$ holds.}

{\bf Lemma 2.3.} $$\xi_fg=\Phi(f,g')$$
The lemma has absolutely the same proof as Lemma 1.3 for the first example.

{\bf Proposition 2.1.} $$\xi_{\{f,g\}}=[\xi_f,\xi_g].$$

{\em Proof.} For any vector field $\xi=\p_{H(Z)}$ we have $$0=(\d\o)(\xi_f,\xi_g,
\xi)=\xi_f\o(\xi_g,\xi)-\xi_g\o(\xi_f,\xi)+\xi\o(\xi_f,\xi_g)$$
$$-\o([\xi_f,\xi_g],\xi)+\o([\xi_f,\xi],\xi_g)-\o([\xi_g,\xi],\xi_f),$$
We have $$\o(\xi_f,\xi)=\tr (H(X)Z-XH(Z))/2=\tr(-H(Z)X+(ZX)')/2$$ $$=-\xi f+
\p(\O_f(\xi)+\tr(ZX)/2),\eqno{(2.4)}$$ $$\o(\xi_g,\xi)=-\xi g+\p(\O_g(\xi)+\tr(ZY)/2),$$
$$\o([\xi_f,\xi],\xi_g)=-\o(\xi_g,[\xi_f,\xi])=[\xi_f,\xi]g-\p(\O_g([\xi_f,\xi]
)+\tr([X,Z]+\xi_fZ-\xi X)Y/2),$$ $$\o([\xi_g,\xi],\xi_f)=[\xi_g,\xi]f-\p(\O_f(
[\xi_g,\xi])+\tr([Y,Z]+\xi_gZ-\xi Y)X/2).$$Now, we have
$$0=-\xi(\xi_fg-\xi_gf)+\xi\o(\xi_f,\xi_g)-\o([\xi_f,\xi_g],\xi)$$ $$
+\p(\xi_f\O_g(\xi)-\xi_g\O_f(\xi)-\O_g([\xi_f,\xi])+\O_f([\xi_g,\xi]))$$
$$+\p\:\tr(\xi_f(ZY)-\xi_g(ZX)-Y\xi_fZ+X\xi_gZ+Y\xi X-X\xi Y$$ $$-
Y[X,Z]+X[Y,Z])/2.$$Using lemma 2.2, we have $$0=-\xi(2\{f,g\}+\p\:\Phi
(f,g))+\xi\{f,g\}-\o([\xi_f,\xi_g],\xi)$$ $$
+\p\:(\xi_f\O_g(\xi)-\xi_g\O_f(\xi)-\O_g([\xi_f,\xi])+\O_f([\xi_g,\xi]))$$
$$+\p\:\tr(Z(\xi_f Y-\xi_g X)+Y\xi X-X\xi Y+2Z[X,Y])/2. $$ Replacing $f$
by $\{f,g\}$ in (2.4), we get $$-\xi\{f,g\}=\o(\xi_{\{f,g\}},\xi)-\p\:
(\O_{\{f,g\}}(\xi)+\tr(ZT/2)) \eqno{(2.5)}$$ where $\xi_{\{f,g\}}=\p
_{H(T)}$. Now, $$\o(\xi_{\{f,g\}}-[\xi_f,\xi_g],\xi)=\p\:(-\xi\Phi(f,g)
-\O_{\{f,g\}}(\xi)-\tr ZT/2)$$ $$
+\p\:(\xi_f\O_g(\xi)-\xi_g\O_f(\xi)-\O_g([\xi_f,\xi])+\O_f([\xi_g,\xi]))$$
$$+\p\:\tr(Z(\xi_f Y-\xi_g X)+Y\xi X-X\xi Y+2Z[X,Y])/2. $$ the fact that
$\o(\xi_{\{f,g\}}-[\xi_f,\xi_g],\xi)$ is an exact derivative for all
$\xi$'s implies that $\xi_{\{f,g\}}-[\xi_f,\xi_g]=0$, as required. 
In particular, $T=[X,Y]+\xi_fY-\xi_gX$.

At the same time, we obtained an identity $$0=\p(-\xi\Phi(f,g)
-\O_{\{f,g\}}(\xi)+\xi_f\O_g(\xi)-\xi_g\O_f(\xi)-\O_g([\xi_f,\xi])+
\O_f([\xi_g,\xi])$$ $$+(Y\xi X-X\xi Y+Z[X,Y])/2).\eqno{(2.6)}$$

{\bf Proposition 2.2.} {\sl The Poisson bracket $\{f,g\}$ satisfies the Jacobi
identity up to an exact derivative term: $$\{\{f,g\},h\}+({\rm cyclic})=
\p\:\Psi(f,g,h)$$ where $\Psi=\d\Phi$ is a trilinear form, with earlier
defined $\Phi$ and  $$(\d\Phi)(f,g,h)=\xi_f
\Phi(g,h)-\Phi(\{f,g\},h)+({\rm cyclic}).$$ } 

{\em Proof}. Let $X=\d f/\d U,~Y=\d g/\d U,~Z=\d h/\d U$. We have 
 $$0=(d\o)(\xi_f,\xi_g,\xi_h)=\xi_f\{g,h\}-\o(\xi_{\{f,g\}},\xi_h)+({\rm
cyclic}).$$With (2.5), this yields $$0=-2\o(\xi_{\{g,h\}},\xi_f)
+\p\:(\O_{\{g,h\}}(\xi_f)+([Y,Z]+\xi_gZ-\xi_hY)X)+({\rm cyclic}).$$
Then, $$2\{\{f,g\},h\}+({\rm cyclic})=  
\p\:(\O_{\{f,g\}}(\xi_h)+([X,Y]Z+Y\xi_hX-X\xi_hY)/2)+({\rm cyclic}).$$
Eliminating the last term with the help of (2.6), we have the same
equation as for the first example; the end of the proof coincides with
that of Proposition 1.2.

Proposition 1.3 holds in the matrix case, too, along with its proof:

{\bf Proposition 2.3.} {\sl The 3-form $\Psi$ satisfies the identity (a
``higher Jacobi'')
$$\sum_{i<j}(-
1)^{i+j}\Psi(\{f_i,f_j\},...,\hat{f}_i,...,\hat{f}_j,...)=0$$ for arbitrary 
4 functions $f_1,f_2,f_3$ and $f_4$.}\\

{\bf References.}\\

[1] Soloviev, V.O., Boundary values as Hamilton variables, {\bf I}, hep-th
9305133, 1993; {\bf II}, q-alg 9501017, 1995

Boundary terms and their Hamiltonian dynamics, hep-th 9601107, 1996

\end{document}